\documentclass[aps,reprint,twocolumn,pra]{revtex4-1}
\usepackage{amsmath,amssymb}
\usepackage{epsfig}
\usepackage{xcolor,datetime}
\usepackage[pdfstartview=FitV,colorlinks=true,citecolor=blue!80,linkcolor=blue!80]{hyperref}

\newcommand{\trace}{\mathrm{tr}}
\newcommand{\dg}{\dagger}
\newcommand{\la}{\lambda}
\newcommand{\mem}{\text{m}}

\newcommand{\at}{\text{a}}
\renewcommand{\a}{\ensuremath{\alpha}}
\renewcommand{\b}{\ensuremath{\beta}}
\newcommand{\s}{\ensuremath{\sigma}}
\renewcommand{\t}{\ensuremath{\tau}}
\newcommand{\Om}{\ensuremath{\Omega_\mem}}
\newcommand{\Oa}{\ensuremath{\Omega_\at}}
\newcommand{\oL}{\ensuremath{\omega_\text{L}}}

\newcommand{\ket}[1]{\ensuremath{\lvert #1 \rangle}}

\newcommand{\mean}[1]{\ensuremath{\langle #1 \rangle}}
\newcommand{\abs}[1]{\ensuremath{\lvert #1 \rvert}}

\begin{document}
\title{Enhancing nanomechanical squeezing by atomic interactions in a hybrid atom-optomechanical system}

\author{Niklas Mann and Michael Thorwart}
\affiliation{I.\ Institut f\"ur Theoretische Physik, Universit\"at Hamburg, Jungiusstra{\ss}e 9, 20355 Hamburg, Germany }

\date{\currenttime, \today}

\begin{abstract}
In a hybrid atom-optomechanical system, the optical coupling of a mechanical mode of a nanomembrane in an optical cavity with a distant interacting atom gas
permits highly non-classical quantum many-body states. We show that the mechanical mode can be squeezed by the back-action of internal excitations of the atoms in the gas. A Bogoliubov approach reveals that these internal excitations form a fluctuating environment of quasi-particle excitations  for the mechanical mode with a gaped spectral density. Nanomechanical squeezing arises due to quasi-particle excitations in the interacting atom gas when the mechanical frequency is close to resonance with the internal atomic transitions. Interestingly, nanomechanical squeezing is enhanced by atom-atom interactions. 
\end{abstract}


\maketitle

In combining the benefits of their different building blocks in a single set-up, hybrid quantum systems are able  to overcome their individual limitations.
For instance, the resolved sideband cooling only allows optical feedback cooling of a nanomechanical oscillator to its quantum mechanical ground state if the oscillator frequency exceeds the photon loss rate in the optical cavity. One promising candidate to overcome this limitation for \textit{low-frequency} oscillators is an atom-optomechanical hybrid system~\cite{Vogell2013,Jockel2015} in which a nanomembrane inside an optical cavity is coupled to a distant cloud of cold $^{87}$Rb atoms placed in the optical potential of the out-coupled light field.
The coupling can be realized in two ways: the membrane can be coupled to either the spatial motion of the atoms in the lattice~\cite{Vogell2013,Jockel2015,Wallquist2009,Hammerer2009,Hunger2011,Zhong2017, Vochezer2018,Mann2018}, or to a transition between the internal states of the atoms~\cite{Vogell2015,Lau2018}.
Apart from membrane cooling, the quantum many-body nature of the atom gas leads to collective atomic motion~\cite{Vochezer2018} and a nonequilibrium quantum phase transition~\cite{Mann2018} from a localized symmetric motional state of the atom cloud to a shifted symmetry-broken state. The 'internal-state coupling' scheme offers the advantage that the frequency scale of internal atomic transitions can be more easily brought into resonance with vibrational frequencies of nanomechanical oscillators. 

Cutting-edge experiments in optomechanics demonstrate coherent state transfer~\cite{Wang2012} and entanglement between the cavity light and the mechanical resonator~\cite{Vitali2007,Genes2008,Asjad2016}, macroscopic quantum coherence~\cite{Liao2016} and squeezed optical~\cite{Purdy2013,Kronwald2014} and mechanical~\cite{Jahne2009,Mari2009,Gu2013,Tan2013,Kronwald2013} states.
It is well known that squeezed states can be generated by nonlinearities~\cite{Kitagawa1993,Wineland1994}, such as, e.g.,  particle-particle interactions in an atomic condensate. 
Also engineering the environment of an optomechanical set-up can produce a squeezed-vacuum reservoir and a transfer of squeezing to the movable mirror may occur~\cite{Gu2013}.
Moreover, spin squeezed states of atoms can be induced by squeezed light~\cite{Sorensen1998,Hald1999,Banerjee1996,Kuzmich1997,Hald2000}, transferring the squeezed state from the light to the atoms.
Combining several of these elements in a single hybrid quantum system may be a key step to generate a robust squeezed mechanical state, that could be largely insusceptible to dissipative effects.

\begin{figure}
\includegraphics[width=\columnwidth]{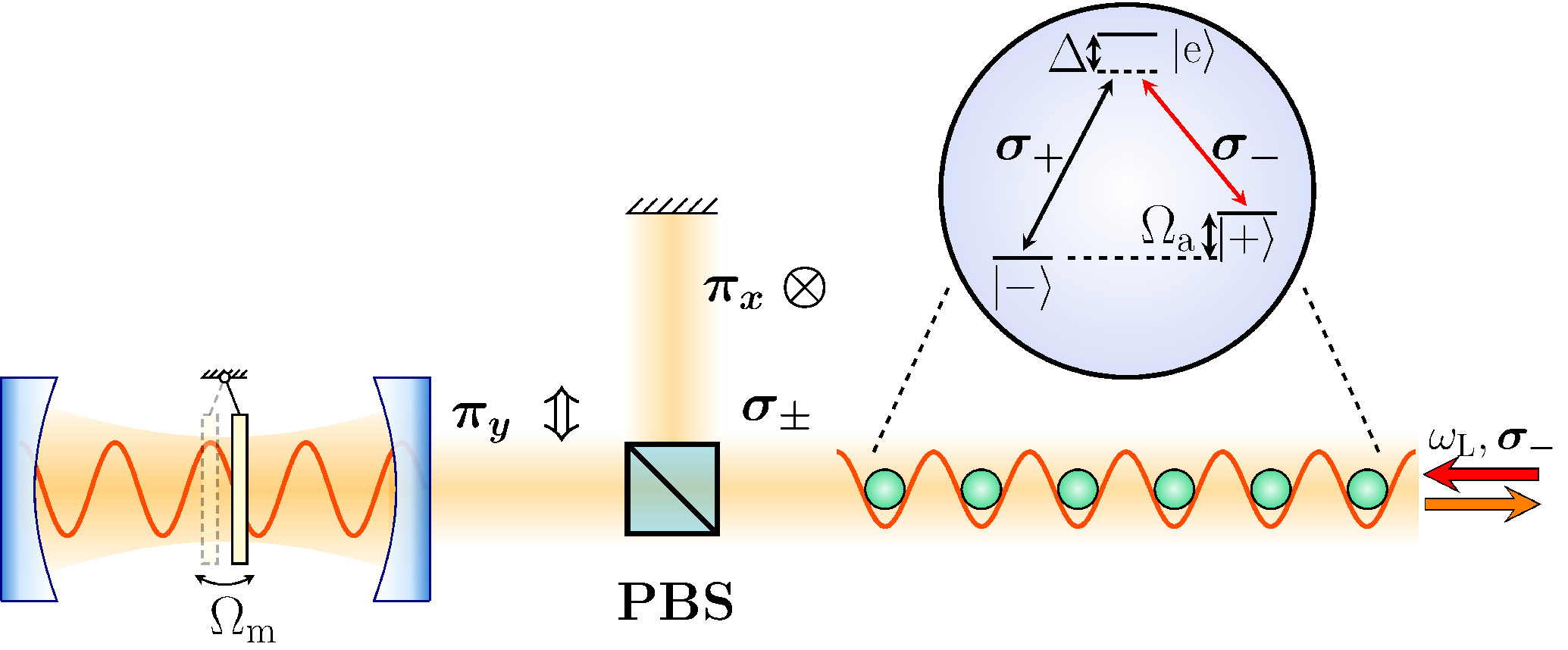}
\caption{Hybrid atom-optomechanical system: A nanomechanical membrane in an optical cavity is coupled to the internal states of a distant atom gas residing in an optical lattice potential.}\label{fig1}
\end{figure}

Inspired by recent experimental progress~\cite{Jockel2015,Zhong2017,Vochezer2018}, we investigate an atom-optomechanical system in the internal-state coupling scheme, see Fig.~\ref{fig1}.
The atoms forming the gas are placed in an optical lattice and two internal electronic states of each atom are addressed. Then, the atom gas is efficiently described by a two-species Bose-Hubbard model.
We assume a weak local atom-atom interaction, and within a Bogoliubov mean-field approximation, the quartic interaction term is reduced to terms quadratic in the bosonic ladder operators. We show that, within this limit and in the presence of the optomechanical coupling to a nanomembrane, the Hamiltonian can be mapped to a system-bath model, where the harmonic bath is realized by the Bogoliubov quasi-particle modes.
We find an analytic expression for the bath spectral density of the quasi-particle modes in the zero-temperature limit, which depends on the atomic interaction strength. Hence, by tuning the atomic interaction, the effective environment of the optomechanical components can be engineered and the quantum state of the nanomechanical oscillator can be controlled. For weak atom-membrane coupling, the system becomes analytically solvable and we have direct access to thermodynamical observables of the membrane and atoms. Most interestingly, 
 the variance of the mechanical displacement coordinate can be reduced to a squeezed nanomechanical state in a wide range of model parameters. In fact, squeezing is enhanced by a finite atom-atom interaction at finite temperatures. 
%
%
\section{Model}
We consider an ensemble of three-level atoms placed in an external optical lattice potential.
The internal states are arranged in a $\Lambda$-type scheme, with the two lowest-energy states being energetically separated by $\Oa$. 
A $\sigma_-$ polarized laser with a finite detuning $\Delta$ drives the transition between the state $\ket{+}$ and the excited state $\ket{\text{e}}$. 
The passing beam is directed onto a polarization beam splitter (PBS) which splits the circularly polarized light into linearly polarized light on two arms of equal length, see  Fig.~\ref{fig1}.
The end of the upper path consists of a fixed mirror and in the left path a nanomembrane is placed inside a low-finesse cavity. We assume a single vibrational mode of the membrane with frequency $\Om$ to be relevant.
The reflected and outcoupled light of both arms is directed back onto the atoms mediating the effective coupling between the atoms and the membrane. 
In a quasistatic picture, a finite displacement of the membrane induces a finite optical phase shift on the propagating beam leading to a rotation of the polarization in the PBS.
The emergent $\sigma_+$ photon can then induce a two-photon transition between the states $\ket{-}\leftrightarrow\ket{+}$ in the atoms, when the resonance condition $\Om\simeq\Oa$ is met.
The back-action of the atoms on the membrane is triggered by a transition of the atoms between the states $\ket{\pm}$, which changes the radiation pressure on the membrane by the emission of an additional $\sigma_+$ photon.

In the bad-cavity limit, the light field and the auxiliary state $\ket{e}$ can be adiabatically eliminated, and the system is effectively described by an extended two-type Bose-Hubbard model ($\hbar=1$)
\begin{align}
H =& \Om b^\dg b + \sum_{k,\s}\epsilon_{k\s} n_{k\s} \nonumber+\frac{U}{2L}\sum_{kpq}\sum_{\s\t}d^\dg_{k-q\s}d^\dg_{p+q\t}d_{p\t}d_{k\s}\\
&-{\la}q_\mem\sum_{k}\left[d_{k+}^\dg d_{k-}+\text{H.c.}\right]\,,\label{eq1}
\end{align}
with $L$ sites and $\s,\t=\pm$. The bosonic creation operator for the mechanical mode is $b^\dg$ and 
$q_\mem=(b+b^\dg)/\sqrt{2\Om}$ is the membrane displacement operator. The bosonic operator $d^\dg_{k\s}$ creates an excitation of the internal state $\s$ of an atom with spatial momentum $k$ and $n_{k\s}=d^\dg_{k\s}d_{k\s}$. The 
atom-atom interaction strength is denoted by $U$, the atom-membrane coupling by $\la$ and the atomic energy dispersion is $\epsilon_{k\s}=\s\Oa/2-2J\cos(k)$, with the hopping amplitude $J$. 
Moreover, we have neglected a term that couples the number of atoms in the $ \s=+$ state to the membrane displacement and a term that represents long-range interaction induced by the light field.
The first is justified by the assumption of weak atom-membrane coupling and low temperature, whereas the latter leads to a redefinition of the quasi-particle mode frequencies that we determine in the following.
The momentum index $k$ runs in equidistant steps of size $2\pi/L$ from $-\pi$ to $\pi$.
For a more detailed derivation of the effective Hamiltonian, see appendix~\ref{app}.

\section{Bogoliubov approach} Assuming the hierarchy $\Om,\,\Oa\gg NU/L,\,\la$ allows us to use a Bogoliubov approximation in the mean field ground state, where all atoms occupy the state with quantum numbers $(k,\s)=(0,-)$. In the Bogoliubov prescription, we associate
\begin{align}\begin{split}
d_{0-}\,,\;d_{0-}^\dg & \rightarrow  \: \sqrt{n_{0-}}\,, \\
d_{0-}^\dg d_{0-}^\dg d_{0-} d_{0-} & \rightarrow \:  n_{0-}(n_{0-}-1)\,, \\
n_{0-} & = N - n_{0+} - \sum_{k\ne 0, \s} n_{k\s}\,.
\end{split}\end{align}
%
%
Via the third equality, the Bogoliubov approach is defined in a particle conserving manner.
For the first line to be acceptable, we require a large number of atoms $N\gg1$ to occupy the ground state.
Then, the approximated Hamiltonian to leading order in $N$ is given by
\begin{align}\begin{split}
H' =& \Om b^\dg b+\Oa' a^\dg a  + \sum_{k\ne 0, \s} E_{k\s}d_{k\s}^\dg d_{k\s}-\sqrt{2N}\la' q_\mem q_\at \\
&  - \la q_\mem \sum_{k\ne 0, \s} d^\dg_{k\s}d_{k\, -\s}+\frac{nU}{2}\sum_{k\ne 0}\left[d_{-k\, -}d_{k\, -}+\text{H.c.}\right]\,,
\end{split}\label{eqq:app-B}\end{align}
with $E_{k\s}=\epsilon_{k\s}-\epsilon_{0-}+nU$ and $\Oa'=\Oa+nU\simeq\Oa$.
Moreover, we have set $a\equiv d_{0+}$, $\la'=\la\sqrt{\Oa}$ and introduced $q_\at=(a+a^\dg)/\sqrt{2\Oa}$ and the particle density $n=N/L$. 
The excited modes ($k\neq0$) in Eq.~\eqref{eqq:app-B} can be diagonalized by introducing quasi-particle modes $c_{k\s}$ with $
d_{k\, -}=\phi_k c_{k\, -} + \theta_k c^\dg_{-k\, -} $
and $c_{k\, +}=d_{k\, +}$.
The coefficients are given by 
\begin{align}\begin{split}
\phi_k =& \sqrt{(\epsilon_{k-}-\epsilon_{0-}+nU+\omega_{k-})/2\omega_{k-}} \, , \\
\theta_k= &  \sqrt{(\epsilon_{k-}-\epsilon_{0-}+nU-\omega_{k-})/2\omega_{k-}} \,,
\end{split}\end{align}
such that bosonic commutation relations are fulfilled since $\phi_k^2-\theta_k^2=1$.
The operators $c_{k\s}$ describe quasi-particle modes that are mixtures of states with positive and negative momenta $\pm k$.
The frequencies are given by 
\begin{align}\begin{split}
\omega_{k-} =& (\epsilon_{k-}-\epsilon_{0-})\sqrt{1+\frac{2nU}{\epsilon_{k-}-\epsilon_{0-}}}\, , \\
\omega_{k+} = &  E_{k+} \,.
\end{split}\end{align}
With these modes, the Hamiltonian transforms to
\begin{align}
H^B&= \Oa a^\dg a + \Om b^\dg b + \sum_{k\ne 0,\s} \omega_{k\s} c^\dg_{k\s}c_{k\s} - \sqrt{2N}\la' q_\mem q_\at\nonumber\\
&-{\la}q_\mem \sum_{k \ne 0}\left[\phi_k c_{k-}^\dg c_{k+} + \theta_k c_{k-} c_{-k+} + \text{H.c.}\right]\,.
\end{align}
%
%
The coupling of the membrane to the quasi-particle modes involves two different processes: One process induces a transition of a quasi-particle between the states $\ket{-}\leftrightarrow\ket{+}$, whose coupling constant scales with $\phi_k^2=(\epsilon_{k-}-\epsilon_{0-}+nU+\omega_{k-})/2\omega_{k-}$. 
The other process involves the annihilation (creation) of two quasi-particles which scales with $\theta_k^2=(\epsilon_{k-}-\epsilon_{0-}+nU-\omega_{k-})/2\omega_{k-}$.
The latter process is induced by particle collisions and is only present for 
a finite $U>0$ as $4\theta_k^2\simeq \sqrt{nU/J}$.

\section{Spectral density of quasi-particle modes}
Under the stated mean-field conditions, the harmonic quasi-particle modes can be integrated out within a Euklidean  path integral formalism. 
For this, we define the interaction Hamiltonian 
\begin{equation}
H_I^B=-{\la} q_\mem \sum_{k\neq0}\left[\phi_k c_{k-}^\dg c_{k+} + \theta_k c_{k-} c_{-k\, +} + \text{H.c.}\right]\,. 
\end{equation} 
We follow Ref.~\cite{Napoli1994} which addresses a similar structure of the coupling Hamiltonian, yet in the context of two-phonon processes in a Caldeira-Leggett model.  
The influence functional in imaginary time, describing the membrane-atom coupling, takes the form
\begin{equation}
\mathcal{F}[\boldsymbol{q}(\cdot)]=Z_\text{Bog}^{-1}\int \mathcal{D}\boldsymbol{x} \exp\left(-S_\text{Bog}[\boldsymbol{x}]-S_I[\boldsymbol{q},\boldsymbol{x}]\right)\,,\label{eq:app-F}
\end{equation}
where $\boldsymbol{x}$ is the $2(L-1)$ component vector that denotes the Bogoliubov modes and with 
$\boldsymbol{q}(\cdot)=(q_1(\cdot), q_2(\cdot))$  denoting mixed coordinates $q_1$ and $q_2$ defined via
\begin{equation}
\begin{pmatrix}
q_1\\
q_2
\end{pmatrix} = \begin{pmatrix}
\cos\chi & -\sin\chi\\
\sin\chi & \cos\chi
\end{pmatrix}
\begin{pmatrix}
q_\mem\\
q_\at
\end{pmatrix}\,,
\end{equation}
with
\begin{align}\begin{split}
\tan\chi=&\Bigl[\Oa^2-\Om^2+\text{sgn}(\Om-\Oa)\\
&\times\sqrt{(\Oa^2-\Om^2)^2+8N\la'^2}\Bigr]/\sqrt{8N}\la'\,.
\end{split}\end{align}
Moreover, $S_\text{Bog}[\boldsymbol{x}]$ is the free atomic Euklidean action excluding the $k=0$ mode and $S_I[\boldsymbol{q},\boldsymbol{x}]$ is the interaction functional, which can be derived from $H_I^B$.
In Ref.~\cite{Napoli1994}, it has been show that if the mean value $\mean{H_I^B}_\beta=0$, with $\mean{O}_\beta=Z^{-1}_\text{Bog}\trace\{Oe^{-\beta\sum_{k\ne 0, \s}\omega_{k\s}c_{k\s}^\dg c_{k\s}}\}$ and the inverse temperature $\beta$, the right-hand side of Eq.~\eqref{eq:app-F} can be drastically simplified by expanding the exponential, i.e., $
\mathcal{F}[\boldsymbol{q}(\cdot)] = \sum_n\frac{1}{n!} \mean{(-S_I[\boldsymbol{q},\boldsymbol{x}])^n}_\beta\,$.
Only terms of even order in $n$ contribute to the sum and reordering the sum leads to the conclusion that the only relevant contribution is the correlator $\mean{H^B_{I}[\boldsymbol{q}(\tau)]H^B_{I}[\boldsymbol{q}(\tau')]}_\beta$.
It follows that the influence action is then simply given by
\begin{equation}
S_\text{infl}[\boldsymbol{q}(\cdot)]=-\int_0^\beta d\tau \int_0^\tau d\tau' \mean{H^B_{I}[\boldsymbol{q}(\tau)]H^B_{I}[\boldsymbol{q}(\tau')]}_\beta\,,
\end{equation}
with the correlator estimated to be
%
%
\begin{align}\begin{split}
\left\langle H^B_{I}[\boldsymbol{q}(\tau)]\right. &\left.H^B_{I}[\boldsymbol{q}(\tau')] \right\rangle_\beta=\sum_{\a,\a'=1,2}\la_\a\la_{\a'}q_\a(\tau)q_{\a'}(\tau')\\
&\times\sum_{k\neq0}\Bigl\{\phi_k^2[n(\omega_{k-})-n(\omega_{k+})]D_{\Delta_k}(\tau-\tau')\\
&+\theta_k^2[n(\omega_{k-})+n(\omega_{k+})+1]D_{\Omega_k}(\tau-\tau')\Bigr\}\,.
\end{split}\end{align}
%
%
Here, we have defined $\Omega_k=\omega_{k-}+\omega_{k+}$, $\Delta_k=\omega_{k+}-\omega_{k-}$, and the free boson propagator in imaginary time $D_\omega(\t)=2 n(\omega)\cosh{\omega\t}+e^{-\omega\t}$.
The coupling constants $\lambda_\alpha$ are given by $\la_1=\la\cos\chi$, $\la_2=\la\sin\chi$ and $n(\omega)=(e^{\beta\omega}-1)^{-1}$ is the Bose-Einstein distribution. 
With this in mind, the influence action results in
%
%
\begin{equation}
S_\text{infl}[\boldsymbol{q}(\cdot)]=\!-\!\!\!\!\sum_{\a,\a'=1,2}\!\!\!\!\la_\a\la_{\a'} \!\!\int_0^{\b} \!\! d\t \!\!\int_0^\t \!\!\! d\eta \, k(\t-\eta)q_\a(\t)q_{\a'}(\eta) \label{eq:Sinfl}\,,
\end{equation}
%
%
with the kernel $k(\t) = \int d\omega \, G(\omega) D_\omega(\t)$, where
\begin{align}\begin{split}
G(\omega)=&\sum_{k\ne 0}\{ \phi_k^2[n(\omega_{k-})-n(\omega_{k+})]\delta(\omega-\Delta_k)\\& + \theta_k^2[n(\omega_{k-})+n(\omega_{k+})+1]\delta(\omega-\Omega_k) \}\,.
\end{split}\end{align}
is the spectral density of the quasi-particle modes.
Most interestingly, in the limit $L,\beta\rightarrow\infty$ with $LNU\ll2\pi^2 J$ the spectral density takes the form
\begin{equation}
G_{T=0}(\omega) \simeq \frac{2L}{\pi} \left(\frac{nU}{4J}\right)^2 \frac{4J}{\abs{\sin\kappa(\omega)}}\frac{1}{(\omega-\Oa-nU)^2}\,,
\end{equation}
where $\kappa(\omega)=2\sin^{-1}\sqrt{(\omega-\Oa-nU)/8J}$. 
It exhibits singularities at $\omega_{\rm min}=\Oa+nU$ and $\omega_{\rm max}=\Oa+nU+8J$, and is only defined between these frequencies.
In the vicinity of the singularities, the spectral density behaves as $G_{T=0}(\omega_{\rm min}+\delta\omega)\sim (\delta\omega)^{-5/2}$ and $G_{T=0}(\omega_{\rm max}-\delta\omega)\sim (\delta\omega)^{-1/2}$.
Moreover, the coupling to the quasi-particle excitations scales with the number of states, i.e., $\sqrt{L}\la$, rather than the number of particles, i.e., $\sqrt{N}\la$, as it is the case for the zero-momentum coupling. 

\begin{figure}
\includegraphics[width=\columnwidth, trim = 20 0 0 0, clip]{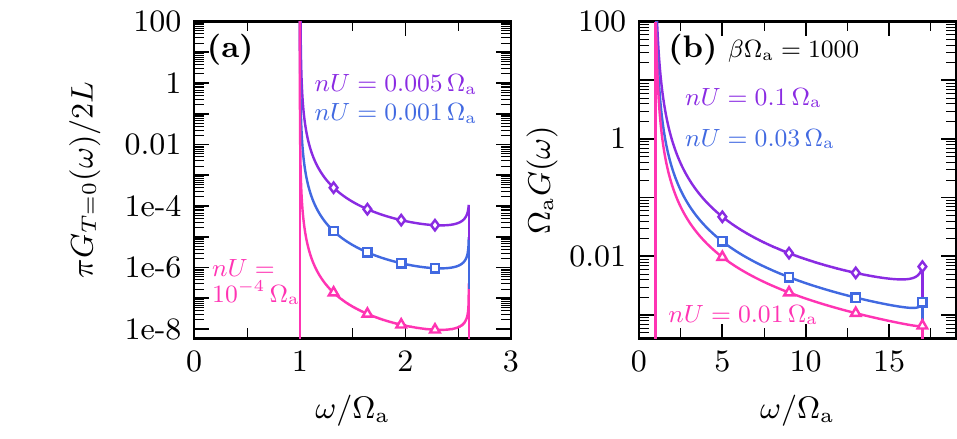}
\caption{(a) Approximated zero-temperature spectral density $G_{T=0}(\omega)$ with $J=0.2\,\Oa$ for different $nU$. (b) Spectral density $G(\omega)$ with $\beta\Oa=1000$, $J=5\,\Oa$ and $L=1000$.}\label{fig2}
\end{figure}
In Fig.~\ref{fig2}, we show $G(\omega)$ for~(a) $\beta\rightarrow\infty$ and~(b) $\beta\Oa=1000$. 
The spectral density has a van Hove-type singularity at the atomic frequency \Oa\ and a second one shifted by 8$J$, which is twice the bandwidth $4J$.
In general, the spectral density scales with the number of modes $L$, meaning the thermodynamic limit $L\rightarrow\infty$ has to be taken with caution.
%
%
\begin{figure*}
\includegraphics[width=\textwidth, trim = 20 0 0 0, clip]{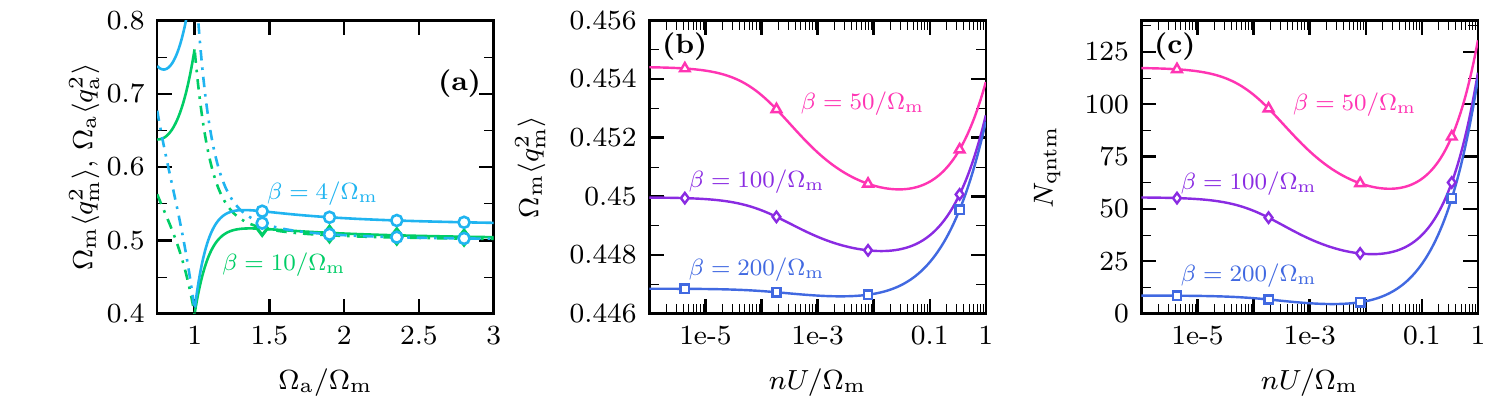}
\caption{(a) Variances of the nanomembrane displacement (solid lines) and the atomic internal state (dashdotted lines) as a function of the atomic frequency $\Oa$ with $nU=0.5\,\Om$. Other parameters chosen are $L=200$, $N=10^6$, $\la=2\times10^{-4}\,\Om$ and $J=0.1\,\Om$.
Moreover, the nanomembrane displacement variance (b) and the number of depleted particles (c) are shown as a function of the interaction strength $nU$.
For (b) and (c), we have chosen $J=0.5\,\Om$, $\Oa=1.04\,\Om$, $N=10^3$, $L=200$ and $\sqrt{2}\la=0.015\,\Om$.
Different colors mark a different inverse temperature as indicated.}\label{fig3}
\end{figure*}
\section{Quantum squeezing of the nanomembrane}
In order to analyze the impact of (weak) atomic interactions on the system state, we study the membrane displacement variance $\mean{q_\mem^2}$.
Hence, we determine the partition function $Z=\int \mathcal{D}\boldsymbol{q}(\cdot) e^{-S_\text{eff}[\boldsymbol{q}]}$ of the reduced system using the influence action of Eq.~\eqref{eq:Sinfl} and the free action of the two harmonic oscillators with frequencies $\Omega_1^2=\Om^2+2\sqrt{2N}\la'\tan\chi$ and $\Omega_2^2=\Oa^2-2\sqrt{2N}\la'\tan\chi$. 
Overall, the reduced system (membrane plus $k=0$ atomic mode) is described by the effective action
\begin{align}\begin{split}
S_\text{eff}[\boldsymbol{q}] =& \frac{1}{2}\sum_\a\int_0^\beta d\t\left[\dot{q}_\a^2(\t)+\Omega_\a^2q_\a^2(\t)\right]\\&-\sum_{\a,\a'}{\la_\a\la_{\a'}}\int_0^{\b} \! d\t\int_0^\t \! d\eta \:k(\t-\eta)q_\a(\t)q_{\a'}(\eta)\:.
\end{split}\end{align}
For a harmonic oscillator, the path integral can be evaluated in a closed analytic expression~\cite{Grabert} by expanding the path $\boldsymbol{q}(\t)=\bar{\boldsymbol{q}}(\t)+\boldsymbol{y}(\t)$ around its extremal path $\bar{\boldsymbol{q}}(\t)$ with deviation $\boldsymbol{y}(\t)$.
The extremal path fulfills the relation $\partial_{\boldsymbol{q}} S_{\text{eff}}[\boldsymbol{q}]\lvert_{\boldsymbol{q}=\bar{\boldsymbol{q}}}=0$.
Via a Fourier expansion of the kernel $k(\t)=\sum_{n=-\infty}^\infty \xi_n e^{i\nu_n\t}/2$ and the deviations $y_\a(\t)=\sum_{n=-\infty}^\infty y_{\a n} e^{i\nu_n\t}/2$, with Matsubara frequencies $\nu_n=2\pi n/\beta$, the effective action takes the form
\begin{align}\begin{split}
S_\text{eff}[\boldsymbol{q}(\cdot)]=&S_\text{eff}[\bar{\boldsymbol{q}}(\cdot)] + \frac{1}{2\beta}\sum_{\a,\a'} 
\sum_{n=-\infty}^\infty y_{\a n} \\
& \times   [\delta_{\a\a'}(\Omega_\a^2+\nu_n^2)-2\la_\a\la_{\a'}\xi_n]y_{\a'n}\,.
\end{split}\end{align}
By performing the integration over the deviations $\boldsymbol{y}(\tau)$, the partition function is found to be
\begin{equation}\label{eq:partition}
Z(\beta) = \frac{\mathcal{N} e^{-S_\text{eff}[\bar{\boldsymbol{q}}]}}{\sqrt{{D}_0}} \prod_{n>0}\frac{1}{{D}_n}\,,
\end{equation}
where $\mathcal{N}$ is a universal normalization constant and the $D_n$ are given by 
\begin{equation}
{D}_n=\prod_{\alpha=1,2}(\Omega_\alpha^2+\nu_n^2)-2\la^2\xi_n(\Omega_1^2\sin^2\chi+\Omega_2^2\cos^2\chi+\nu_n^2)\,.
\end{equation}
Finally, the nanomembrane displacement variance is estimated via the first derivative of the logarithm of the partition function $\mean{q_\mem^2}=-(\beta\Om)^{-1}\partial_{\Om} \text{ln} Z $.

In Fig.~\ref{fig3}(a), we show the variance of the nanomembrane displacement (solid) in dependence of the atomic transition frequency $\Oa$. In the vicinity of the resonance $\Om \lesssim \Oa$, the variance is strongly suppressed due to two-mode squeezing, occurring when $\mean{q_\mem^2} \Om \le 1/2$. 
For this set of parameters, the effect of the quasi-particle excitation is negligibly small due the scaling with $\sim \sqrt{L} \la$ of the coupling. Moreover, the atomic internal state variance $\mean{q_\at^2}$  (dashdotted lines) 
describes a squeezed state when $\mean{q_\at^2} \Oa \le 1/2$. It shows the opposite behavior, such that for $\Oa \lesssim \Om$, both constituents exchange their role, and the atomic quantum many-body state is strongly squeezed in the vicinity of the resonance. Hence, we can conclude that squeezing of the nanomembrane displacement only occurs when quasiparticle excitations are possible above the spectral gap. 

Interestingly, by enhancing the coupling to the excited quasiparticle modes, finite atom-atom interaction may have a positive effect for achieving a squeezed nanomechanical state.
This is shown in Fig.~\ref{fig3}(b) for different temperatures $1/\beta$.
By increasing the atom-atom interaction $nU$, the variance $\mean{q_\mem^2}$ of the nanomembrane displacement first decreases and then grows again.
The position and magnitude of the minimal variance (or, the maximal squeezing) depends on the temperature and is influenced by two processes: 
First, the weak interaction limit is dominated by differences of the thermal excitation of the states with same momentum, but opposite internal state, leading to a scaling with $\sim n(\omega_{k-})-n(\omega_{k+})$ in the spectral density.
Second, for rather strong atomic interaction compared to the thermal energy scale, the dominant part are excitations of the Bogoliubov quasiparticles induced by the interaction. This process scales with $\sim(nU)^2[ n(\omega_{k-})+n(\omega_{k+})+1]$.

Due to the optomechanical coupling, the squeezing of the nanomembrane variance is also reflected in the number 
of atoms $N_\text{qntm}=\sum_{k\ne 0, \s} \mean{d_{k\s}^\dg d_{k\s}}$ excited from the mean-field ground state. 
As shown in Fig.~\ref{fig3}(c), squeezing is maximal when the depletion  of the atomic mean-field  ground state is minimal.

Although these results are limited to weak atom-atom interactions, they give clear evidence that atom interactions can actually be used to enhance squeezing of an external degree of freedom at finite temperature, in this case a nanomechanical mode. Further studies beyond the Bogoliubov approach may not only bring new insight for the present hybrid atom-optomechanical system, but may also define a new route for exploring internal-state coupling schemes in Bose-Einstein condensate-cavity systems \cite{Hemmerich1,Hemmerich2,Ritsch}.

\section{Experimental realization} Current experimental set-ups rely on the external-state coupling scheme~\cite{Jockel2015,Zhong2017,Vochezer2018}. 
Yet, the observation of the shown results is possible by measuring the power spectrum of the membrane displacement, 
even in these set-ups. 
Rather than a transition between two internal atomic states, the atom-membrane coupling induces transitions 
between motional bands of the atoms with different band index. 
In order to achieve two-mode squeezing with a reduced variance of the membrane displacement, an atomic transition frequency slightly larger than the membrane frequency is favorable.
For the external-state coupling scheme, a membrane with a frequency of several hundreds of kilohertz translates to lattice depths of the order of 2000 recoil energies. 
This leads to a strong atom confinement in the corresponding direction and by tuning the perpendicular confinement potential, the effective one-dimensional atom-atom interaction can be tuned.

\section{Conclusions} We have investigated the effect of finite atom-atom interactions in a two-component atom gas on the nanomechanical vibrational state of a distant nanomembrane. 
In the zero-temperature limit, the spectral density of the quasi-particle excitations induced by atomic interactions can be determined within a Bogoliubov approach. 
The spectral density is gaped and exhibits a two peak structure with a dominant peak at the atomic transition frequency and a reduced peak at the largest attainable frequency in the reduced atomic system. We show that even 
at finite temperature, the quasi-particle excitations can be used to enhance the two-mode squeezing of displacement  variance of the nanomembrane, thereby creating  highly nontrivial quantum many-body states involving a squeezed nanomechanical mode and an interacting atom gas coupled in a hybrid optomechanical set-up. Squeezing of the nanomembrance displacement only occurs when quasiparticle excitations are possible above the spectral gap. 

\begin{acknowledgements} This work was supported by the Deutsche Forschungsgemeinschaft. We thank M. Ueda for fruitful discussions and for the kind hospitality at the University of Tokyo where parts of this work have been carried out (N.M.). We thank the MIN Graduate School and the Department of International Affairs of the Universit\"at Hamburg for financial support of this stay.
We thank M.\ Reza Bakhtiari for useful discussions. 
\end{acknowledgements}

\appendix
\section{Effective Hamiltonian of the Hybrid System}\label{app}
In order to obtain an effective Hamiltonian of the hybrid system, we start from 
the total system Hamiltonian consisting of five parts, i.e.,
\begin{equation}
H_\text{tot} = H_\mem + H_\at + H_\text{l} + H_\text{m-l} + H_\text{a-l}\,. \label{eq:app-Hlin}
\end{equation}
Each of the first three terms describes one of the three compounds: the nanomembrane, the atomic condensate and the light field, respectively.
The coupling between the light field and the membrane or the atoms is described by the last two terms.
The nanomembrane is modeled as a single vibrational mode with 
\begin{equation}
H_\mem = \Om b^\dg b\,.
\end{equation} 
The atomic part is a condensate of bosonic atoms which have three relevant internal states marked by $\s\in\{-,+,e\}$. The Hamiltonian is given by 
\begin{equation}
H_\at = \sum_\s \int dz \Psi_\s^\dg(z)\left[V_\s(z) -\omega_R \partial_z^2 + \frac{g}{2}n(z)\right]
\Psi_{\s}(z)\,,
\end{equation}
with the atomic particle density $n(z)=\sum_\s \Psi_\s^\dg(z)\Psi_\s(z)$. With $\oL$ being the laser frequency and $m$ the mass of the atoms, the recoil frequency is $\omega_R=\oL^2/2m$. Moreover, the strength of the one-dimensional atom-atom interaction is $g$ and the potential $V_\s(z)$ includes the energy of the corresponding internal state and an optical lattice potential for the external spatial atomic degree of freedom.  
The three internal atomic states are energetically arranged as shown in Fig.\ref{fig1}. 
The light modes have two possible optical polarizations, which are described by the operators $a_\omega, b_\omega$ and included over a bandwidth $2\theta$ around $\oL$. They are described by
\begin{equation}
H_\text{l}= \int_{\oL-\theta}^{\oL+\theta} d\omega \Delta_\omega\left[a_\omega^\dg a_\omega + b_\omega^\dg b_\omega\right]\,,
\end{equation}
with $\Delta_\omega=\omega-\oL$.
All of the above defined operators fulfill bosonic commutation relations.

An external laser with field strength $\a$ and polarization $\s_-$ is included by the linear replacement
\begin{equation}
a_\omega \rightarrow a_\omega + \delta(\omega-\oL)e^{-i\oL t}\a\,,
\end{equation}
such that the coupling between the light field and the membrane (atoms) can be linearized.
In a reference frame rotating with the laser frequency, the linearized membrane-light field interaction takes the form
\begin{equation}
H_\text{m-l}=\la_\mem q_\mem\int \frac{d\omega}{\sqrt{2\pi}}\left(a_\omega+a^\dg_\omega+b_\omega+b^\dg_\omega\right)\,,
\end{equation}
with the coupling strength $\la_\mem$.
After elimination of the auxiliary state $\sigma={e}$, the atom-light field coupling 
\begin{align}\begin{split}
H_\text{a-l}&=\int\frac{d\omega}{\sqrt{2\pi}}\int dz \sin(z)\sin(\tfrac{\omega}{\oL}z)\left[\la_-b_\omega\Psi_+^\dg(z)\Psi_-(z)\right.\\&\left.+\la_-b_\omega^\dg\Psi_-^\dg(z)\Psi_+(z)+\la_+(a_\omega+a_\omega^\dg)\Psi_+^\dg(z)\Psi_+(z)\right] \, 
\end{split}\end{align}
consists of several terms describing different effects. 
The first two terms lead to transitions between the atomic internal states under the creation (or annihilation) of a polarized photon.
On the other hand, the last term couples the $\s=+$ internal states to the photon field quadrature, in a similar manner as in the external coupling scheme~\cite{Vogell2013}.
In assuming low temperatures $\beta\Oa\gg1$, this last term can be consequently neglected for our following approach.

In order to obtain an effective Hamiltonian that describes the atom-membrane coupling directly, we adiabatically eliminate the light modes.
We start with the linearized Hamiltonian~\eqref{eq:app-Hlin} in the interaction picture with respect to $U(t)=\exp\left\{i\int d\omega \,\Delta_\omega[a_\omega^\dg a_\omega + b_\omega^\dg b_\omega]t\right\}$.
The formal solution of the Schr\"odinger equation in the interaction picture reads
\begin{equation}
\ket{\psi(t)_I}=\mathcal{T} \exp\left\{-i\int_0^t dt' H(t')_I\right\}\ket{\psi(0)}\,,
\end{equation}
with the time-ordering operator $\mathcal{T}$ and the index $I$ indicating the interaction picture.
Next, we expand the right-hand side for small time increments $\delta t$. Up to second order, the relevant terms read
\begin{align}\begin{split}
\ket{\psi(\delta t)_I} \simeq& \left\{1 -i \int_0^{\delta t} dt H(t)_I\right.
\\& \left.- \int_0^{\delta t} dt \int_0^{t} dt' H(t)_I H(t')_I \right\}\ket{\psi(0)}\,.
\end{split}\end{align}
Moreover, we assume that the initial state is a product state $\ket{\psi(0)}=\ket{\psi}_{\text{a+m}}\otimes\ket{\text{vac}}_\text{l}$, where $\ket{\text{vac}}_\text{l}$ is the vacuum state of the light field and $\ket{\psi}_\text{a+m}$ is an arbitrary state in the atom-membrane subspace.
Hence, we assume $a_\omega\ket{\psi(0)}=b_\omega\ket{\psi(0)}=0$ and define noise-increment operators
\begin{align}
\delta A(t) =& \int_t^{t+\delta t} ds \int \frac{d\omega}{\sqrt{2\pi}}a_\omega(s)_I\,,\\
\delta B(t) =& \int_t^{t+\delta t} ds \int \frac{d\omega}{\sqrt{2\pi}}b_\omega(s)_I\,,\\
\delta C(t) =& \int_t^{t+\delta t} ds \int \frac{d\omega}{\sqrt{2\pi}} \sin(\tfrac{\omega}{\oL}z) b_\omega(s)_I\,.
\end{align}
Next, we take the limit $\delta t\rightarrow0$ and assume that the noise-increment operators after a time step $\delta t$ do not depend on their form at the earlier time, which is equivalent to the Markov approximation. Then, we can derive a quantum stochastic Schr\"odinger equation in the Ito form with $d\ket{\psi(t)} = \ket{\psi(t+dt)}-\ket{\psi(t)}$.
The differential noise operators, e.g., $\delta A(t)\rightarrow_{\delta t\rightarrow0} dA(t)$, follow the Ito rules
\begin{align}
dA(t)dA^\dg(t)=dB(t)dB^\dg(t)=dt\,,\\
dB(t)dC^\dg(t)=dC(t)dB^\dg(t)=\sin (z) dt\,,\\
dC(t)dC^\dg(t)=\sin (z) \sin (z') dt\,.
\end{align}
Equipped with these relations, the effective Hamiltonian 
\begin{align}
H_\text{eff} =& \Om b^\dg b + \sum_{\s=\pm}\int dz \Psi^\dg_\s\left[V_\s-\omega_R\partial_z^2+\frac{g}{2}n(z)\right]\Psi_\s(z)\nonumber\\
&-\frac{ \lambda_-\lambda_{\rm m} }{4} q_\mem\int dz \sin(2z)\left[\Psi_+^\dg(z)\Psi_-(z)+\text{H.c.}\right] 
\end{align}
is found, where we have neglected a term that describes long-range interaction induced by the light field.
The last term describes the effective atom-membrane coupling.
For the following, we assume an optical lattice of the form $V_\s(z)=\s\Oa/2+V\sin(2z)/2$ and make the ansatz $\Psi_\s(z)=\sum_j w(z-z_j)d_{j\s}$ for the atom field operators with the Wannier function $w(z)$, the lattice sites $z_j=j\pi+\pi/4$ and bosonic ladder operators $d_{j\s}$.
With this ansatz and after a Fourier transformation $d_{k\s}=\sum_jd_{j\s}e^{ijk}/\sqrt{L}$, we arrive at the Hamiltonian given in Eq.\ \eqref{eq1} of the main text.
The hopping rate is defined via
\begin{equation}
J=\int dz w^*(z-z_i)\left[\omega_R\partial_z^2-\frac{V}{2}\sin(2z)\right]w(z-z_{i+1})\,,
\end{equation}
the local interaction strength
\begin{equation}
U=g\int dz \abs{w(z)}^4\,,
\end{equation}
and the effective atom-membrane coupling
\begin{equation}
\la = \frac{ \lambda_-\lambda_{\rm m} }{4} \int dz \sin(2z)\abs{w(z)}^2\,.
\end{equation}

\end{document}